\documentclass[pra,twocolumn,aps,showpacs,nofootinbib]{revtex4-1}

\usepackage{physics}
\usepackage{amsmath}
\usepackage{amssymb}
\usepackage{graphicx}
\usepackage{float}
\graphicspath{{PNG/}}
\usepackage{xcolor}
\usepackage[utf8]{inputenc}

\usepackage{lipsum}

\makeatletter

\newcommand*\bigcdot{\mathpalette\bigcdot@{.5}}
\newcommand*\bigcdot@[2]{\mathbin{\vcenter{\hbox{\scalebox{#2}{$\m@th#1\bullet$}}}}}
\makeatother

\begin{document}
\title{Arbitrary interaction quench phenomena in harmonically trapped two-body systems}

\author{A. D. Kerin}
\author{A. M. Martin}

\affiliation{School of Physics, University of Melbourne, Parkville, VIC 3010, Australia}

\date{\today}

\begin{abstract}

We consider the evolution of two contact-interacting harmonically-trapped particles following an arbitrary quench in interaction strength. We calculate the post-quench particle separation as a function of time and the total post-quench energy. When quenching from any non-zero interaction strength to zero interaction strength we observe that the total energy and particle separation diverge. In particular, the divergent behaviour arises \textit{always} and \textit{exclusively} when quenching \textit{to} the non-interacting regime. The source of the divergence is a power-law tail in the probability distribution of particle separation. This validates and builds upon previous work that found divergent behaviour arises when quenching from the strongly interacting limit to the non-interacting limit in both the two and three-body cases.
 
\end{abstract}
\maketitle

\section{Introduction}
\label{sec:Intro}

Understanding the non-equilibrium behaviour of quantum systems is strongly relevant to a variety of very fundamental problems such as how quantum systems equilibrate \cite{eisert2015quantum} and quantum thermodynamics more generally \cite{polkovnikov2011colloquium}. Ultracold atomic gases are very fertile ground for studying quantum thermodynamics \cite{PhysRevLett.102.160401, liu2010three, Nature463_2010, PhysRevLett.107.030601, Science335_2010,  Cui2012, PhysRevA.85.033634, PhysRevA.86.053631, cetina2015decoherence, levinsen2017universality, colussi2018dynamics, colussi2019bunching, bougas2021few,  bougas2022dynamical}. A high degree of experimental control is possible over such systems \cite{phillips1982laser, chu1991laser, tannoudji1992atom, PhysRevLett.96.030401, grunzweig2010near, serwane2011deterministic, zurn2012fermionization, wenz2013few, zurn2013pairing, murmann2015two, kaufman2021quantum} and they exist in a regime where various analytical techniques are highly applicable e.g. the Fermi pseudo-potential \cite{fermi1936motion, huang1957quantum}.

We consider two particles in a three-dimensional isotropic harmonic trap interacting via a contact interaction. We are interested in the behaviour of the system after a sudden change, a quench, in the interaction strength from one arbitrary value to another. We use known analytic wavefunctions \cite{busch1998two} to calculate the total energy and particle separation evolution after the quench. Previous calculations of quenched two and three-body systems find that the system size diverges after a quench from very strong interactions (unitarity) to no interactions \cite{kerin2020two, kerin2022quench, kerin2022effects}. However, it is unclear how having a finite non-zero interaction strength will affect the system.

Additionally, we must note that the physical system of interest, two contact-interacting bodies in a three-dimensional harmonic trap quenched in interaction strength, can be realised with current experimental capabilities. Methods to construct few-atom systems are well understood \cite{PhysRevLett.96.030401, grunzweig2010near, serwane2011deterministic, zurn2012fermionization, wenz2013few, zurn2013pairing, murmann2015two} and exploiting Feshbach resonance is a well-known reliable method of controlling the interaction strength \cite{fano1935feshbackh, feshbach1958feshbackh, tiesinga1993feshbackh, chin2010feshbach}. In particular the evolution of particle separation of a quenched system has been experimentally measured \cite{guan2019density}. However, that experiment considered a quench in trap geometry rather than interaction strength as we consider here.

This work is structured in the following way. In Sec. \ref{sec:Overview} we review the interacting two-body wavefunction first derived by Ref. \cite{busch1998two}. In Sec. \ref{sec:Quench} we calculate the energy expectation, $\langle E \rangle$, of the post-quench state to determine if the system size diverges. In particular, we examine quenches between the unitary/non-interacting limits and arbitrary interaction strengths and quenches between two arbitrary interaction strengths. In Sec. \ref{sec:ParticleSepEvo} we present calculations of the particle separation evolution.

\section{Overview of the Two-Body Problem}
\label{sec:Overview}

In this work we consider two distinguishable particles in a three-dimensional isotropic harmonic trap interacting via a contact interaction. We use the Fermi-Huang pseudopotential \cite{fermi1936motion, huang1957quantum} to describe the interaction. The centre-of-mass (COM) motion separates out as a simple harmonic oscillator Hamiltonian with mass $M=m_{1}+m_{2}$ and position $\vec{R}=(m_{1}\vec{r}_{1}+m_{2}\vec{r}_{2})/M$. Where $m_{\rm i}$ and $\vec{r}_{\rm i}$ are the mass and position of the i\textsuperscript{th} particle respectively. However, the relative motion is more complicated due to the interaction term. The relative Hamiltonian is given
\begin{eqnarray}
\hat{H}_{\rm rel}&=&-\frac{\hbar^2}{2\mu}\nabla^{2}_{r}+\frac{\mu\omega^2r^2}{2}
+\frac{2\pi\hbar^2 a_{s}}{\mu}\delta^3(r)\frac{\partial}{\partial r}(r \bigcdot),
\label{eq:Hamiltonian}
\end{eqnarray}
where $\hat{H}_{\rm total}=\hat{H}_{\rm COM}+\hat{H}_{\rm rel}$, $\vec{r}=\vec{r}_{1}-\vec{r}_{2}$,  $\mu=m_{1}m_{2}/M$, $\omega$ is the trapping frequency, and $a_{s}$ is the $s$-wave scattering length which characterises the interaction. The dot inside the brackets of the derivative is to indicate that the derivative acts on $r\psi$ (where $\psi$ is the wavefunction) not just $r$.

\begin{figure}
\includegraphics[height=5cm, width=8cm]{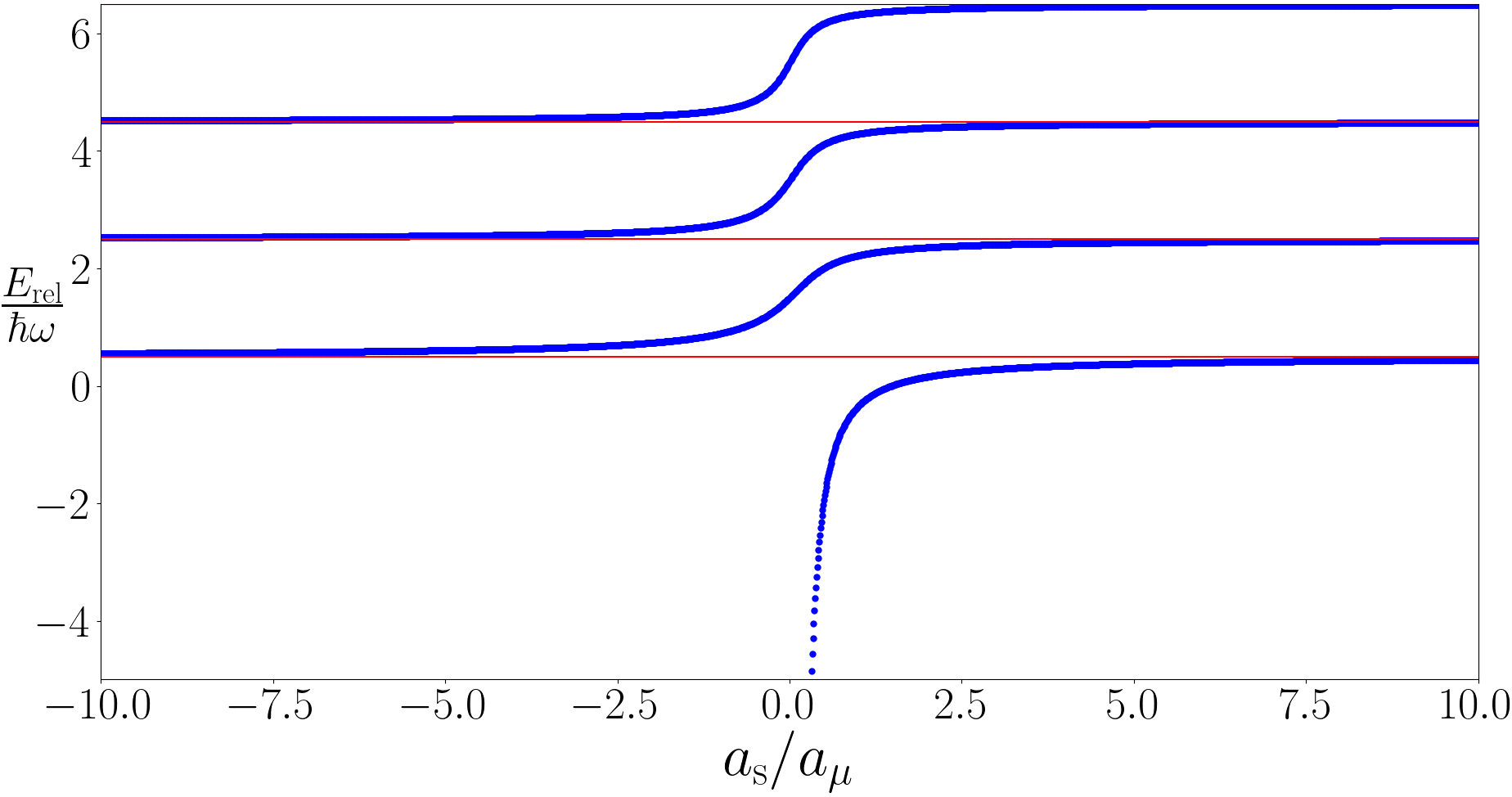}
\caption{The energy of the interacting two-body wavefunction, Eq. (\ref{eq:TwoBodyWavefunction}), as a function of $s$-wave scattering length. The red horizontal lines correspond to $E_{\rm rel}=(2n+1/2)\hbar\omega$, the energies in the unitary limit.}
\label{fig:EnergySpec}
\end{figure}

The eigenfunctions of Eq. (\ref{eq:Hamiltonian}), i.e. the interacting wavefunctions, were first found by Ref. \cite{busch1998two} and are given
\begin{eqnarray}
\psi_{\nu} (r)&=&N_{\nu}\Gamma(-\nu)e^{r^2/2 a^2_{\mu}}U\left(-\nu,\frac{3}{2},\frac{r^2}{a_{\mu}^2}\right),\label{eq:TwoBodyWavefunction}\\
N_{\nu}^{-1} &=& \sqrt{ 2\pi^2 a_{\mu}^3 \frac{\Gamma(1-\nu)\left[\psi^{(0)}\left(-\nu-\frac{1}{2}\right)-\psi^{(0)}(-\nu) \right]}{\nu\Gamma\left(-\nu-\frac{1}{2}\right)} }.\nonumber\\\label{eq:TwoBodyWavefunctionNorm}
\end{eqnarray}
where $a_{\mu}=\sqrt{\hbar/\mu\omega}$, and $\nu$ is the energy pseudo quantum number with $E_{\rm rel}=(2\nu+3/2)\hbar\omega$. The values of $\nu$ are obtained by solving the transcendental equation
\begin{eqnarray}
\frac{a_{\mu}}{a_{s}} = \frac{2\Gamma(-\nu)}{\Gamma\left( -\nu - \frac{1}{2} \right)}.
\end{eqnarray}
The energy spectrum is presented in Fig. \ref{fig:EnergySpec}. Note the bound state present for small positive $a_{s}$.

\section{Quench Dynamics}
\label{sec:Quench}

In general a quench is a sudden change in some variable of the system, e.g. a sudden change in $\omega$ is a quench in trapping frequency. Here we are concerned with the effects of a quench in $a_{s}$. In experiment such quenches have been implemented  \cite{cetina2016ultrafast} by exploiting Feshbach resonance \cite{feshbach1958feshbackh, chin2010feshbach}. 

The wavefunction of the system, $\psi(t)$, as a function of time after the quench can be written
\begin{eqnarray}
\psi(t) &=& \sum_{j=0}^{\infty} \bra{\phi_{j}}\ket{\psi(0)}\phi_{j}e^{-iE_{j}t/\hbar},\label{eq:TimeDepPsi}
\end{eqnarray}
where $t=0$ is the time of the quench, $\psi(0)$ is the pre-quench wavefunction, $\phi_{j}$ are the eigenstates of the post-quench system, and $E_{j}$ are the associated eigenenergies. The overlap terms, $\bra{\phi_{j}}\ket{\psi(0)}$, are presented in Ref. \cite{kerin2020two}. Quenches in $a_s$ only change the relative Hamiltonian not the COM Hamiltonian. As such, only the relative part of the total wavefunction is affected. The COM wavefunction simply integrates to one in all calculations, and we do not need to consider it further.

When quenching from the strongly interacting limit to the non-interacting limit the expectation of the particle separation, $\langle r(t) \rangle$, diverges \cite{kerin2020two}. This divergence is also present in the three-body case \cite{kerin2022quench, kerin2022effects}. This divergence arises because a $1/r^2$ tail forms in the probability distribution of particle separation when $t\neq n\pi/\omega$. The first moment of this distribution, $\langle r(t) \rangle$, is poorly defined due to the $1/r$ tail of the integrand. In the three-body case we can only consider quenches between the unitary and non-interacting regimes, but in the two-body case we can consider arbitrary quenches. This naturally leads one to ask if the divergence is present for other quenches and, if so, under what circumstances.

\subsection{Presence of the Divergence}
\label{sec:Divergence}

It is not immediately obvious whether the divergence is unique to the unitarity to non-interacting limit (backwards) quench or if there are other quenches with the divergence present. Rather than calculate $\langle r(t) \rangle$ for a variety of quenches it is more efficient to calculate $\langle E \rangle$,
\begin{eqnarray}
\langle E \rangle &=& \bra{\psi(t)}\hat{H}\ket{\psi(t)},\nonumber\\
 &=&\sum_{j=0}^{\infty}E_{j}|\bra{\psi(0)}\ket{\phi_{j}}|^2.\label{eq:EnergyExpecation} 
\end{eqnarray}
If $\langle r(t) \rangle$ is divergent, then so is the potential energy and the total energy $\langle E \rangle$. 

\begin{figure}[H]
\center
\includegraphics[height=5cm, width=8cm]{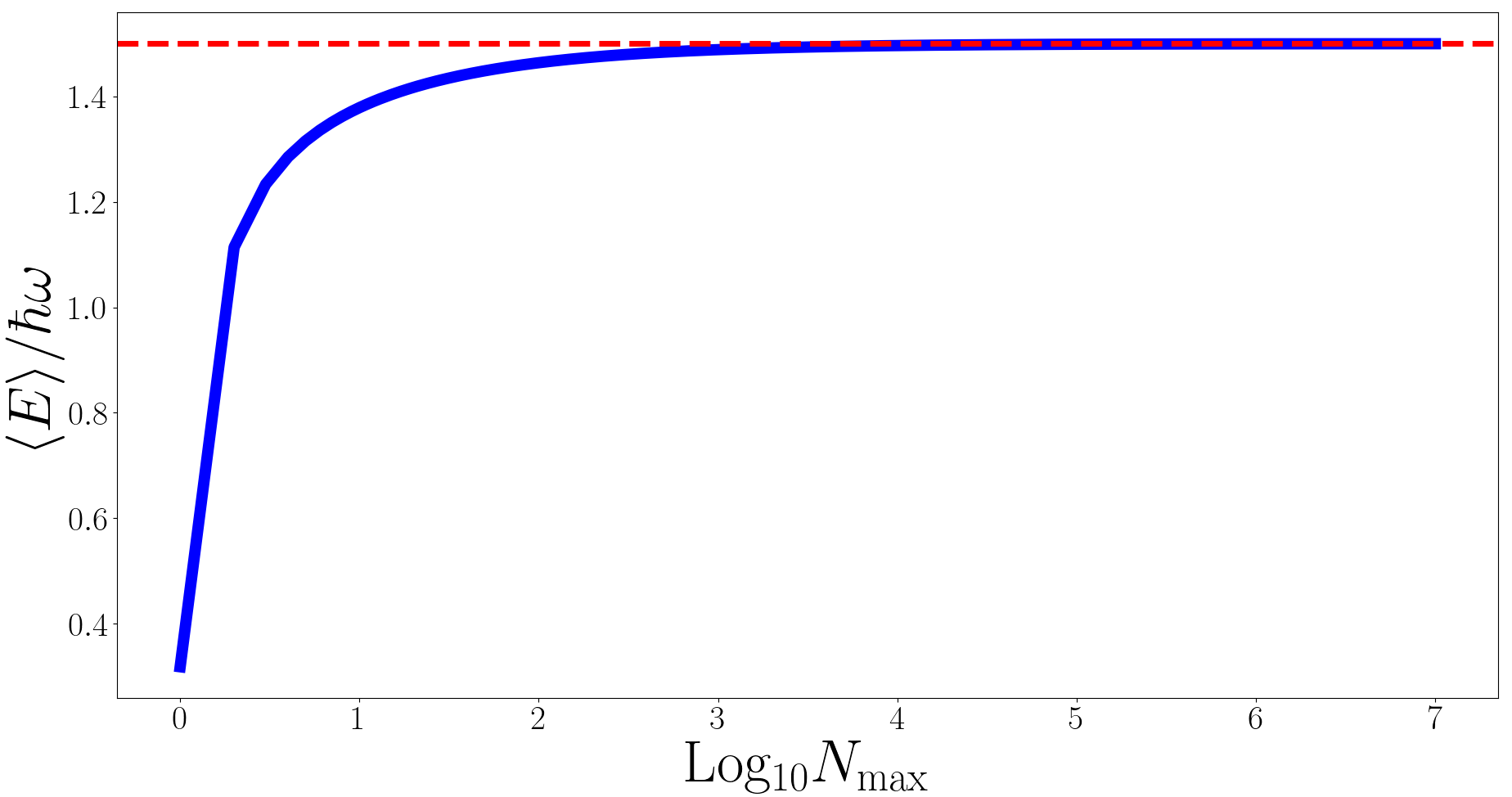}
\includegraphics[height=5cm, width=8cm]{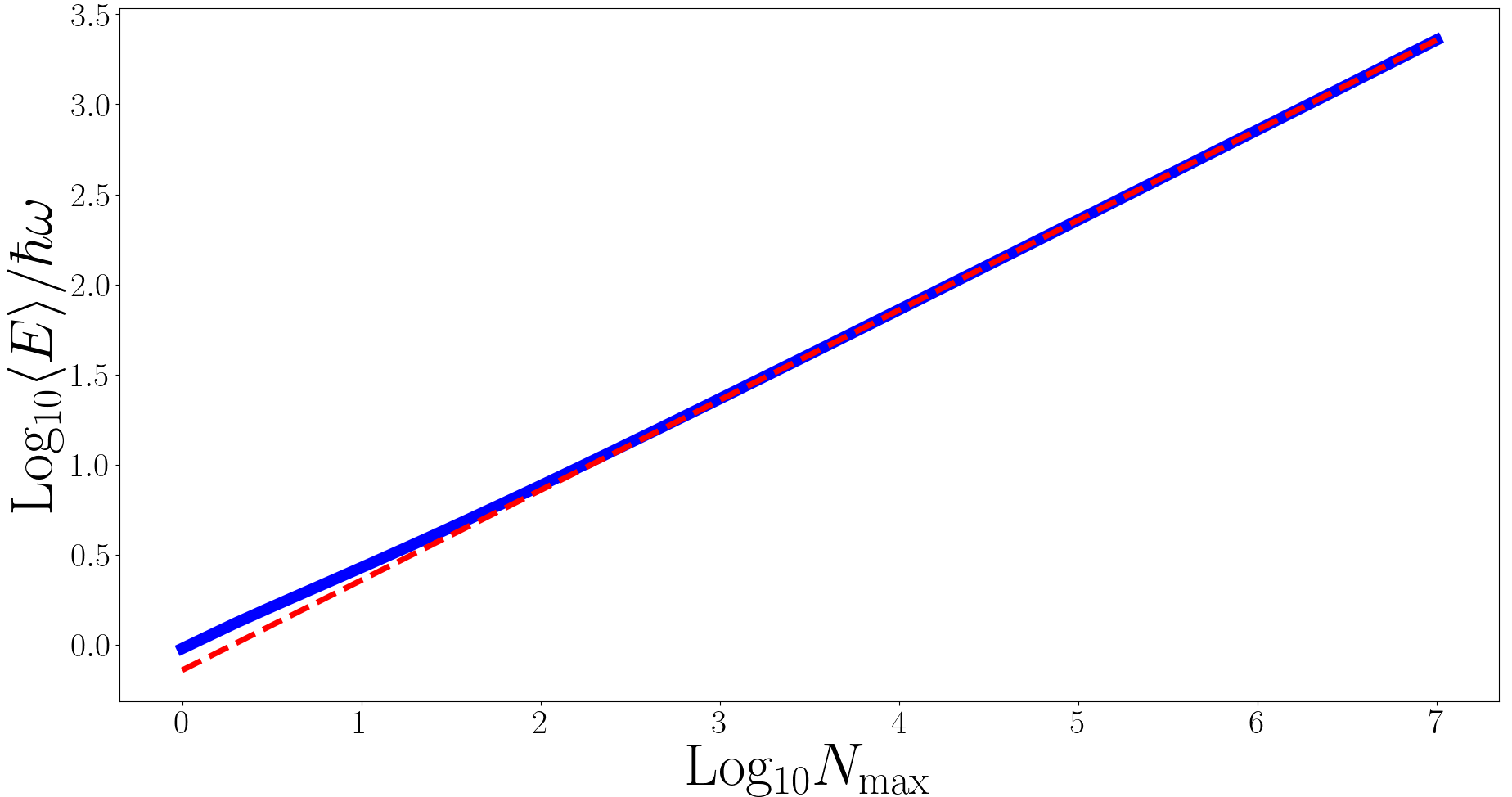}
\caption{The post-quench energy expectation as a function of the number of terms in the expansion in Eq. (\ref{eq:EnergyExpecation}), $N_{\rm max}$. (Upper panel) $\langle E \rangle$ for the forwards quench. The dashed red line indicates $\langle E \rangle=1.5\hbar\omega$. (Lower panel) $\langle E \rangle$ for the backwards quench. The dashed red line indicates $\langle E \rangle=0.7 \sqrt{N_{\rm max}}\hbar\omega$. In both cases the initial state is the ground state. }
\label{fig:LimitExpectE}
\end{figure}

In Fig. \ref{fig:LimitExpectE} we display $\langle E \rangle$ against the number of terms we evaluate up to in Eq. (\ref{eq:EnergyExpecation}), $N_{\rm max}$, for the forwards (non-interacting limit to unitarity) and backwards quenches. It is clear that the forwards quench is convergent and the backwards is divergent with $\langle E \rangle \approx 0.7 \sqrt{N_{\rm max}}\hbar\omega$ for $N_{\rm max}\gtrsim 100$. This is consistent with previous results \cite{kerin2020two}.

Next we turn our attention to quenches involving arbitrary scattering length. There are two cases here. There are quenches between two arbitrary scattering lengths, and there are quenches between the unitary/non-interacting limits and arbitrary $a_s$. We begin with the former case. 

\begin{figure}
\center
\includegraphics[height=6cm, width=7cm]{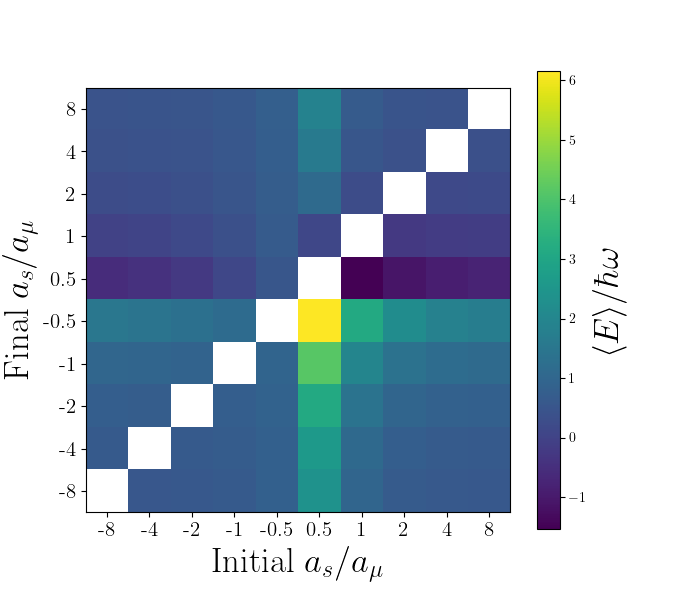}
\caption{$\langle E \rangle$ for a variety of quenches between arbitrary $s$-wave scattering lengths. Closer to yellow indicates larger $\langle E \rangle$ and closer to blue is smaller. In all cases the initial state is the ground state. The blank spaces along the main diagonal are when the initial and final scattering lengths are the same and are therefore not quenches. }
\label{fig:EHeatMap}
\end{figure}

In Fig. \ref{fig:EHeatMap} we plot $\langle E \rangle$ for quenches between two arbitrary scattering lengths where the initial state is the ground state. In all cases we find that $\langle E \rangle$ converges. $\langle E \rangle$ peaks when quenching from small positive $a_s$ to small negative $a_s$. In this case the initial state is a bound state, and it is being projected onto a basis without bound states. There are significant contributions from higher order terms which have large energies which results in large $\langle E \rangle$. These higher order contributions mean that systems initially in a bound state have relatively large $\langle E \rangle$. This is why $\langle E \rangle$ increases sharply as initial $a_{s}$ moves from negative to positive. There is also a sharp decrease in $\langle E \rangle$ when the final $a_{s}$ crosses from negative to positive. This is due to contributions from the bound state which has large negative energy which decreases $\langle E \rangle$.

\begin{figure}
\center
\includegraphics[height=4.5cm, width=8cm]{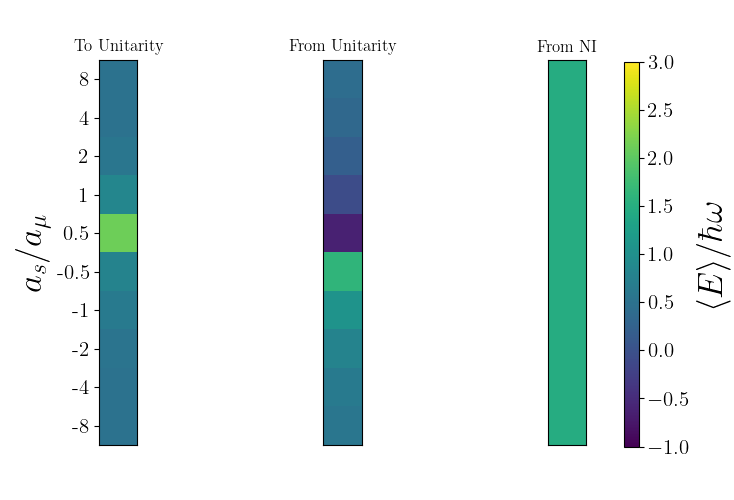}
\includegraphics[height=4.5cm, width=8cm]{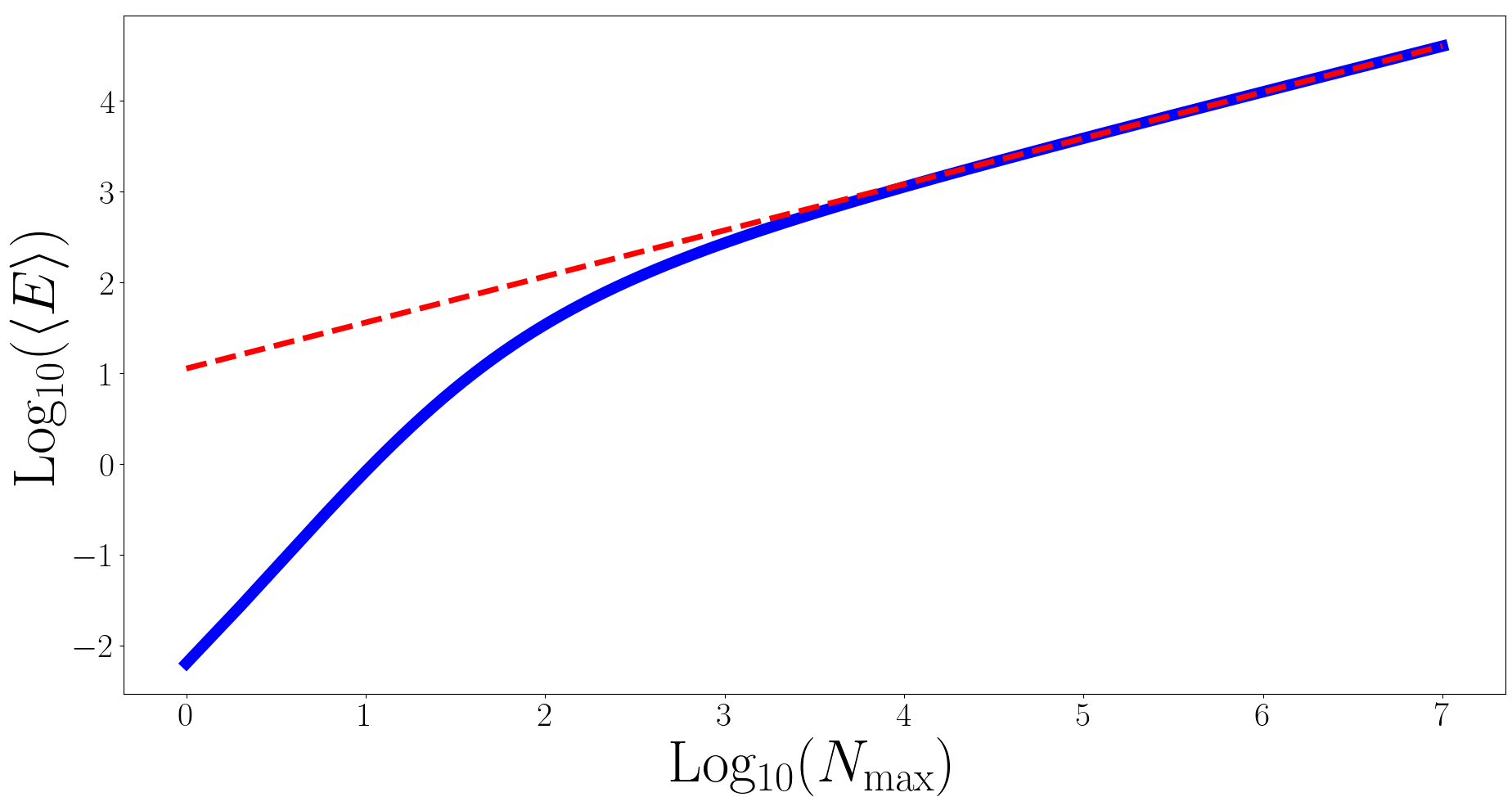}
\caption{Post-quench $\langle E \rangle$ for quenches between the non-interacting (NI) or unitary limits and arbitrary values of $a_{s}$. (Upper panel) $\langle E \rangle $ for quenches between a variety of $s$-wave scattering lengths and unitarity and for quenches from the non-interacting regime to arbitrary $a_{s}$. (Lower panel) $\langle E \rangle$ for a quench from $a_{s}=0.1a_{\mu}$ to the non-interacting regime as a function of $N_{\rm max}$. The red dashed line corresponds to $\langle E \rangle =11.2\sqrt{N_{\rm max}}\hbar\omega$. In all cases the initial state is the ground state.}
\label{fig:LimitArbE}
\end{figure}

Now we turn to quenches between the non-interacting/unitary limits and finite $a_{s}$. In the upper panel of Fig. \ref{fig:LimitArbE} we present $\langle E \rangle$ following quenches between arbitrary $s$-wave scattering length and unitarity and from the non-interacting regime to arbitrary $a_s$. For these quenches we find that $\langle E \rangle$ converges whatever the value of $a_s$. For the \textit{towards} unitarity quenches $\langle E \rangle$ peaks near initial $a_{s}=0$. Small initial $a_{s}$ corresponds to values of $\nu$ which differ significantly from those of the unitary spectrum. When projecting the initial weakly interacting wavefunction onto the unitary basis there are significant contributions from higher order terms, because the unitary states are a poor basis for the weakly interacting states. Higher order terms have larger energies hence $\langle E \rangle$ peaks near initial $a_{s}=0$. For the quenches \textit{from} unitarity $\langle E \rangle$ is strongly influenced by the energy of the post-quench ground state. The initial state (the pre-quench ground state) overlaps strongly with the ground state of the final system. For small positive $a_{s}$ $\langle E \rangle$ is large and negative because of the bound state with large negative energy and for small negative $\langle E \rangle$ is relatively large and positive because the corresponding ground state has relatively large positive energy. In contrast quenches \textit{from} the non-interacting regime do not cause $\langle E \rangle$ to change from its pre-quench value. 

In the lower panel of Fig. \ref{fig:LimitArbE} we plot $\langle E \rangle$ against $N_{\rm max}$ for a quench from $a_{s}=0.1a_{\mu}$ to the non-interacting limit. For this quench $\langle E \rangle$ diverges. In fact, a quench from any non-zero $s$-wave scattering length to the non-interacting regime results in a divergence in $\langle E \rangle$. This is not entirely unexpected. After all, the interacting wavefunction for arbitrary $a_s$ has the same functional form as the unitary wavefunction. Importantly, this demonstrates that the divergence behaviour is not confined to the backwards quench.

\section{Dynamic Sweep Theorem}

The results of Sec. \ref{sec:Divergence} are consistent with the dynamic sweep theorem \cite{tan2008large}. The dynamic sweep theorem relates change in energy after a quench in scattering length and/or trap geometry to the initial conditions of the system and the properties of the quench. If the trap geometry does not change it can be written
\begin{eqnarray}
\frac{dE}{dt} &=& \frac{\hbar^2\Omega C(t)}{8\pi \mu}\frac{d[-a_{s}(t)^{-1}]}{dt}, \label{eq:DynamicSweepTheorem1}
\end{eqnarray}
where $\Omega C(t)$ is the is the integrated contact intensity of the system. It is related to the probability of two particles being in the same place at the same time. For an instantaneous quench Eq. (\ref{eq:DynamicSweepTheorem1}) becomes
\begin{eqnarray}
E_{\rm f} = E_{\rm i} + \frac{\hbar^2 \Omega C(0)}{8\pi \mu}\left( \frac{1}{a_{\rm i}} - \frac{1}{a_{\rm f}} \right), \label{eq:DynamicSweepTheorem}
\end{eqnarray}
where $E_{\rm i}$ and $E_{\rm f}$ are the initial and final energies and $a_{\rm i}$ and $a_{\rm f}$ are the initial and final scattering lengths. In the static interacting two-body case it is given \cite{enss2022complex}
\begin{eqnarray}
\Omega C = 16\pi^2 \lim_{r\rightarrow0}|r\psi_{\nu}(r)|^2 = 16\pi^3 a_{\mu}^2 N_{\nu}^2,
\end{eqnarray}
and in the static non-interacting case $\Omega C=0$. The predictions of the dynamic sweep theorem  match our calculations of $\langle E \rangle$. That is, Eqs. (\ref{eq:EnergyExpecation}) and (\ref{eq:DynamicSweepTheorem}) are in good agreement.  It should be noted that $\Omega C(0)/a_{\rm i}$ goes to zero in the limit of $a_{\rm i}\rightarrow 0$. This means that $E_{\rm f}=E_{\rm i}$ for quenches \textit{from} the non-interacting regime, which is again consistent with previous results.

In prior work it was suggested that the divergent behaviour may be caused by the instantaneous nature of the quench and/or the zero-range nature of the interaction \cite{kerin2020two, kerin2022quench, kerin2022effects}. Using the dynamic sweep theorem we can show that the divergence is indeed a result of the instantaneous quench. Consider a quench where the scattering length changes as
\begin{eqnarray}
a_{s}(t) &=& \frac{a_{\rm f}-a_{\rm i}}{\pi}\arctan(x t) + \frac{a_{\rm f}+a_{\rm i}}{2},
\end{eqnarray}
where $x$ is a parameter to describe the speed of the quench. We assume that the quench happens on a timescale faster than the dynamical response time such that $C(t)=C(0)$ and that $a_{s}$ changes only between $t=-t'$ to $t=t'$. From Eq. (\ref{eq:DynamicSweepTheorem1}) we obtain
\begin{eqnarray}
&& E_{\rm f} - E_{\rm i} =\nonumber\\
&& \frac{\hbar^2\Omega C(0)}{\mu} \left[ \frac{(a_{\rm f}-a_{\rm i})\arctan(xt')}{\pi^2(a_{\rm f}+a_{\rm i})^2 -4(a_{\rm f}-a_{\rm i})^2\arctan(xt')^2}  \right].\nonumber\\\label{eq:FiniteQuenchNoDivTheorem}
\end{eqnarray}
For a quench to the non-interacting regime Eq. (\ref{eq:FiniteQuenchNoDivTheorem}) will only diverge in the limit of an instantaneous quench ($x \rightarrow \infty$). A contact-interacting system can be quenched to the non-interacting regime and not diverge provided the quench is not instantaneous. This demonstrates that the divergent behaviour is a result of instantaneous quenches.

\section{Particle Separation Evolution}
\label{sec:ParticleSepEvo}

In Sec. \ref{sec:Divergence} we demonstrate that the system displays divergent behaviour when quenching to the non-interacting regime from an interacting system (either unitarity or finite $a_s$) by calculating $\langle E \rangle$ for a variety of quenches. However, $\langle E \rangle$ is difficult to experimentally measure and techniques to measure the positions of individual atoms have been successfully implemented in experiment\cite{bergschneider2018spin, guan2019density}. This prompts us to look more closely at the effects of the quench on the particle's positions. The probability distribution of $r$ post-quench is
\begin{eqnarray}
P(r,t) &=& \sum_{j,k=0}^{\infty} \bra{\psi(0)}\ket{\phi_{j}}\bra{\phi_{k}}\ket{\psi(0)}\nonumber\\
&&\times 4\pi r^2\phi_{j}(r)\phi_{k}(r)e^{-i(E_{k}-E_{j})t/\hbar},\label{eq:ProbDistrib}
\end{eqnarray}
and the expectation value of $r$ is
\begin{eqnarray}
\langle r(t) \rangle &=& \sum_{j,k=0}^{\infty} \bra{\psi(0)}\ket{\phi_{j}}\bra{\phi_{k}}\ket{\psi(0)}\nonumber\\
&&\times \bra{\phi_{j}}r\ket{\phi_{k}}e^{-i(E_{k}-E_{j})t/\hbar}. \label{eq:SepExpect}
\end{eqnarray}
The cross terms, $\bra{\phi_{j}}r\ket{\phi_{k}}$, are presented in Ref. \cite{kerin2020two}.

\begin{figure}
\center
\includegraphics[height=5cm, width=8cm]{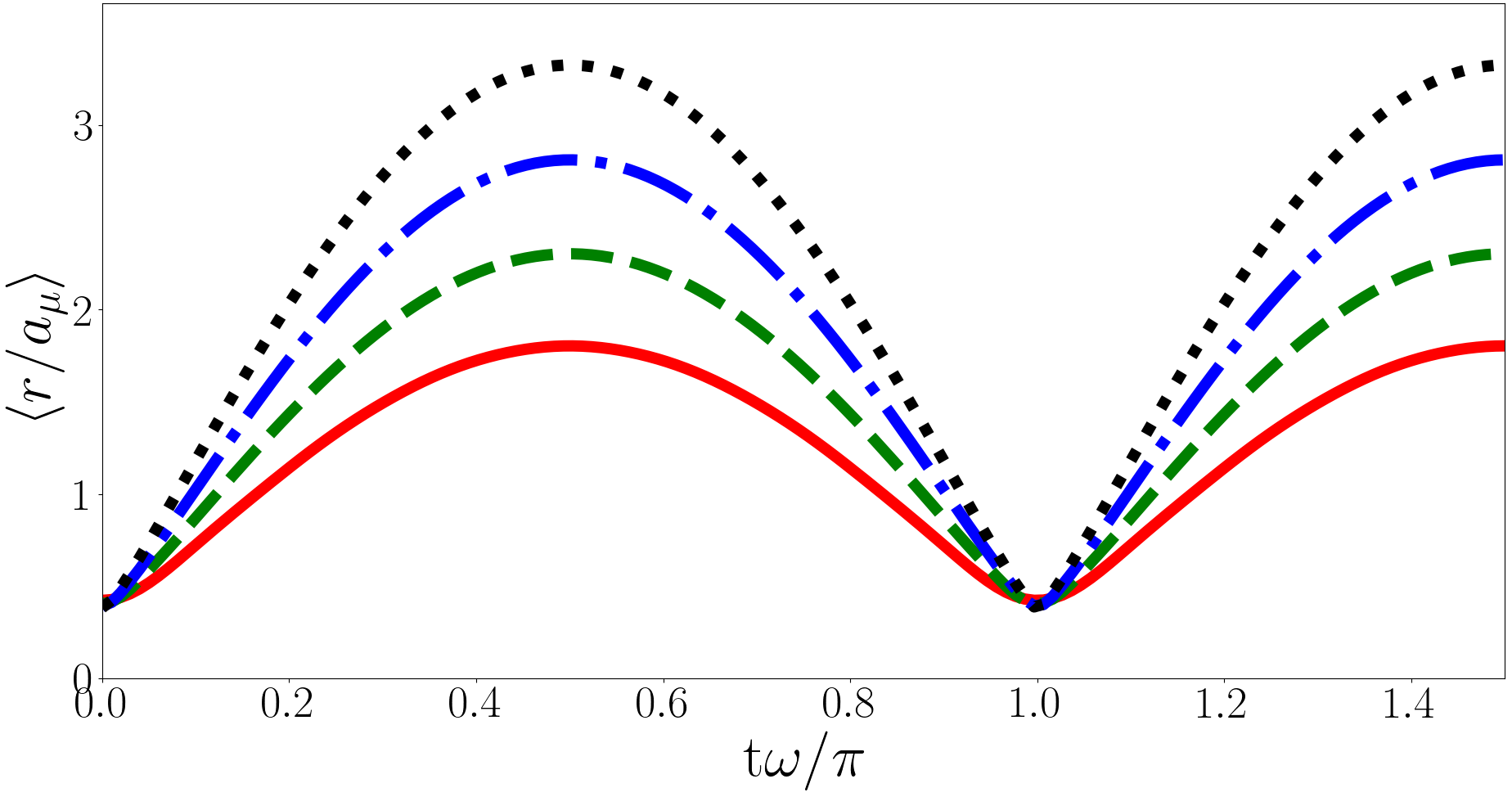}
\includegraphics[height=5cm, width=8cm]{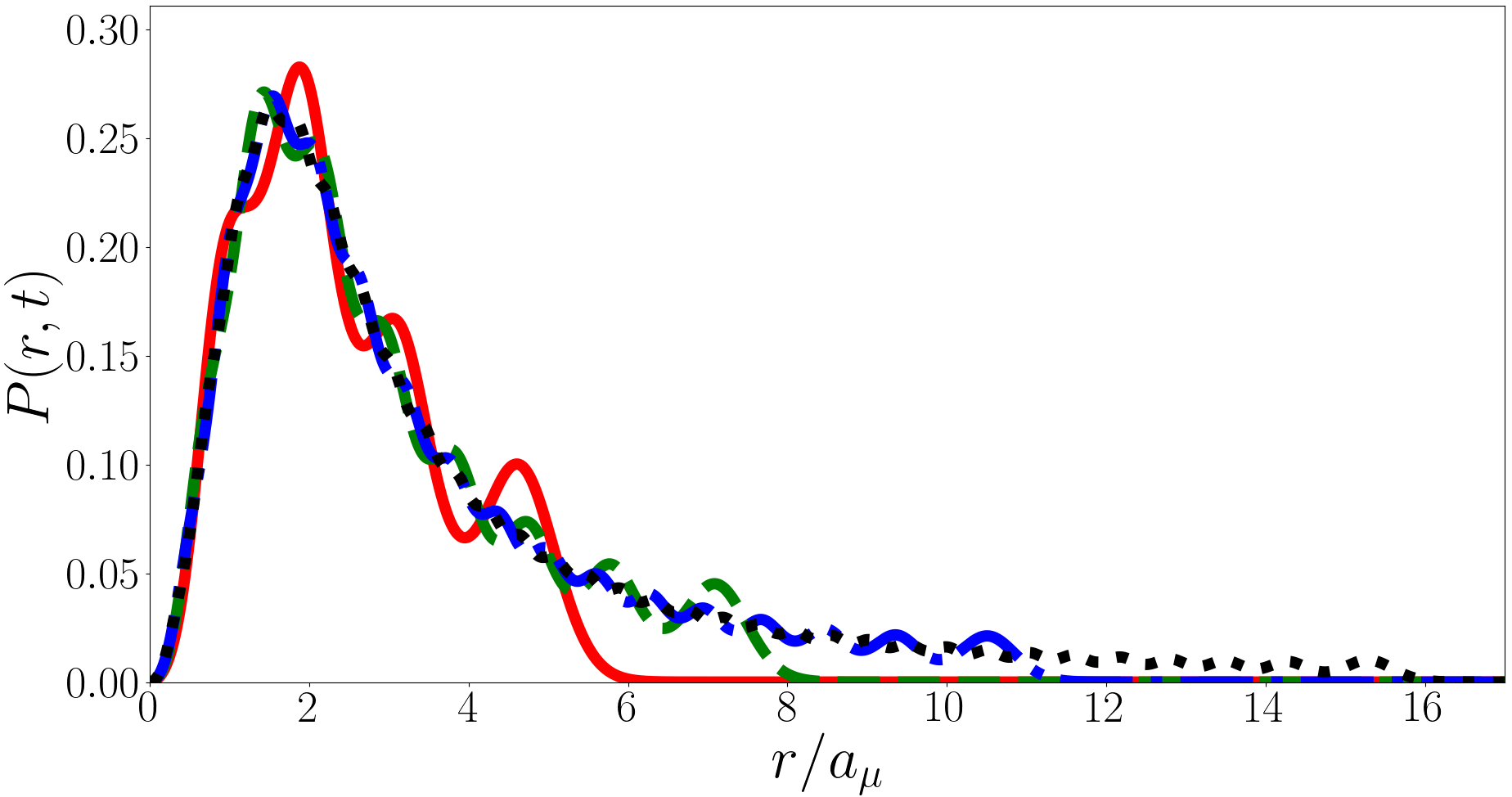}
\caption{Particle separation for the $a_s=a_{\mu}$ to non-interacting quench. (Upper panel) $\langle r(t) \rangle$ for various values of $N_{\rm max}$. The divergence is clearly apparent. (Lower panel) $P(r,t=\pi/2\omega)$ for various values of $N_{\rm max}$. Note the long $1/r^2$ tail and the exponential cut-off moving right as $N_{\rm max}$ increases. In both cases the solid red line corresponds to $N_{\rm max}=8$, the dashed green line to $N_{\rm max}=16$, the dot-dashed blue line to $N_{\rm max}=32$, and the dotted black line to $N_{\rm max}=64$.}
\label{fig:ArbNIDivergence}
\end{figure}

In Fig. \ref{fig:ArbNIDivergence} we plot Eqs. (\ref{eq:ProbDistrib}) and (\ref{eq:SepExpect}) evaluated for the $a_s=a_{\mu}$ to non-interacting quench for various summation cut-offs, $N_{\rm max}$. In line with our results above we observe that $\langle r(t) \rangle$ diverges. Similar to the reverse quench there is a long $1/r^2$ tail in $P(r,t)$ away from $t=n\pi/\omega$. For finite $N_{\rm max}$ the tail ends in an exponential cut-off which disappears as $N_{\rm max}\rightarrow \infty$. This tail is the source of the divergence as it causes the first moment of the distribution to be poorly defined \cite{kerin2020two}.

\begin{figure}[H]
\center
\includegraphics[height=5cm, width=8cm]{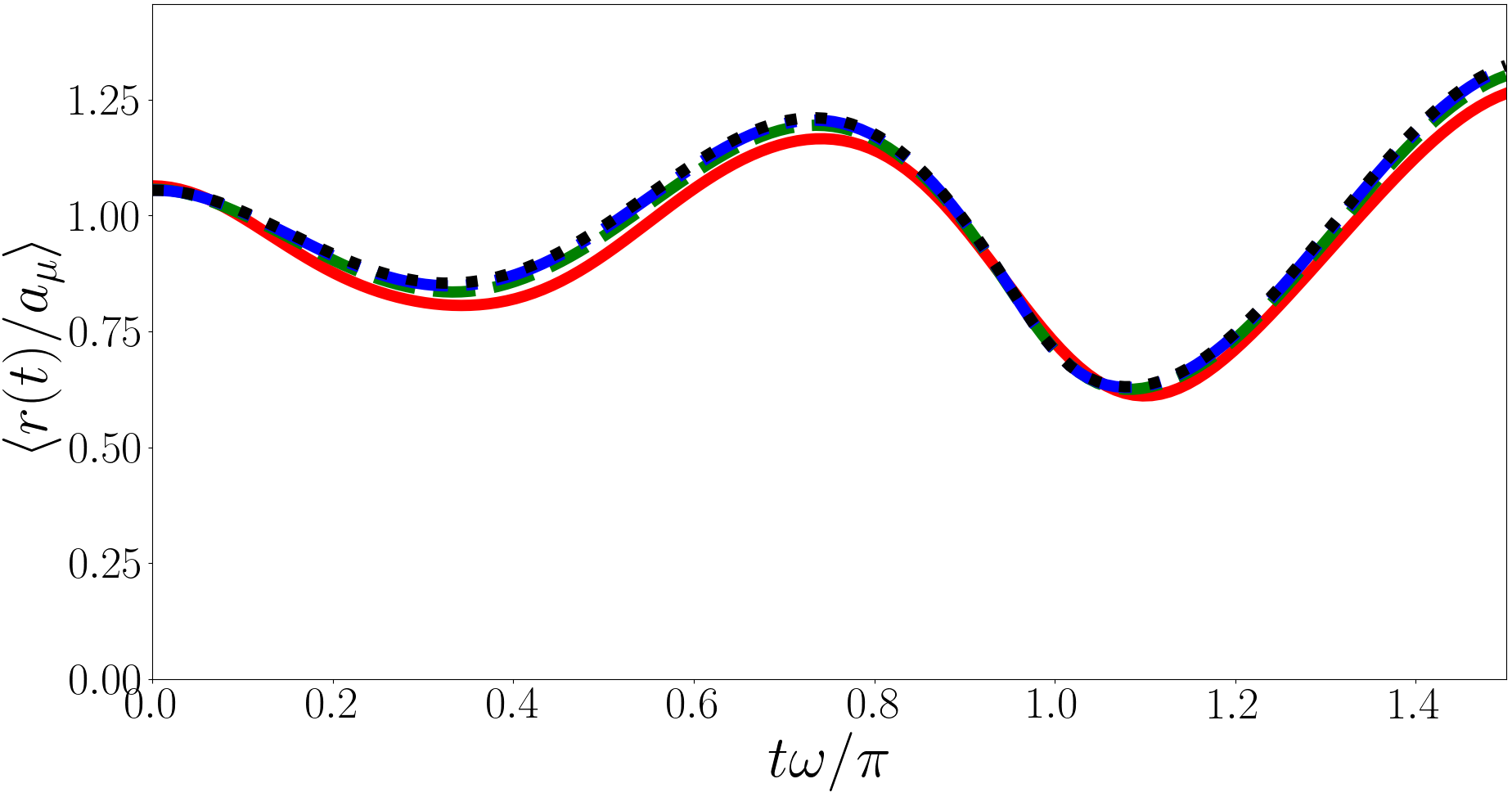}
\includegraphics[height=6cm, width=8cm]{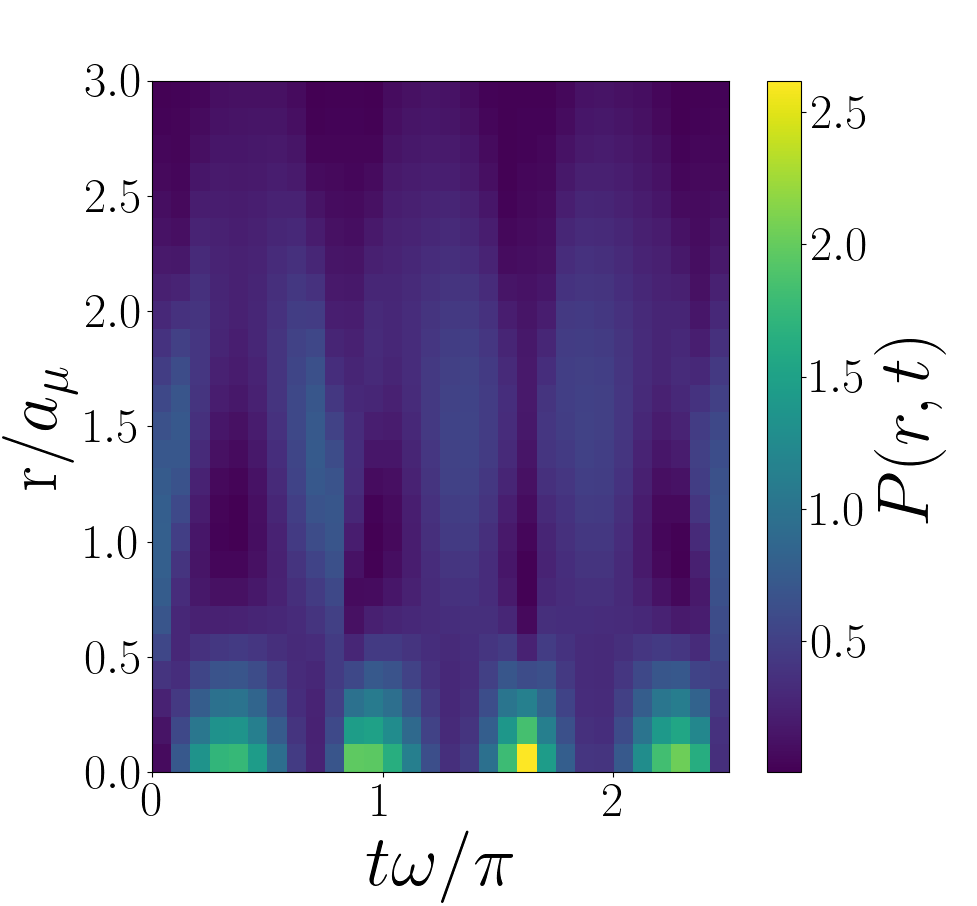}
\caption{ The evolution of particle separation over time for a quench from $a_{s}=-0.1a_{\mu}$ to $a_{s}=a_{\mu}$. (Upper panel) $\langle r(t) \rangle$ for various values of $N_{\rm max}$. The solid red line corresponds to $N_{\rm max}=4$, the dashed green line to $N_{\rm max}=8$, the dot-dashed blue line to $N_{\rm max}=16$, and the dotted black line to $N_{\rm max}=32$. (Lower panel) $P(r,t)$ for the same quench as the upper panel. Dark blue indicates low probability density and yellow indicates high probability density. $P(r,t)$ was evaluated for $N_{\rm max}=64$.}
\label{fig:ParticleSepEvo}
\end{figure}

In the quenches to the unitary/non-interacting limits $P(r,t)$, and therefore $\langle r(t) \rangle$, oscillate periodically \cite{kerin2020two}. This is because in those cases the frequencies of each term in Eqs. (\ref{eq:ProbDistrib}) and (\ref{eq:SepExpect}) are even multiples of $\omega$. If we quench to a finite value of $a_{s}$ then $E_{k}-E_{j}$ is no longer an integer multiple of $\omega$. This leads to the terms in Eqs. (\ref{eq:ProbDistrib}) and (\ref{eq:SepExpect}) oscillating at frequencies which are not multiples of one another which results in aperiodic behaviour. In Fig. \ref{fig:ParticleSepEvo} we plot Eqs. (\ref{eq:ProbDistrib}) and (\ref{eq:SepExpect}) for a quench from $a_{s}=-0.1a_{\mu}$ to $a_{s}=a_{\mu}$, and the aperiodic behaviour is evident. The interaction is initially repulsive, and after the quench the interaction is attractive with the system supporting a bound state. This is reflected in the evolution of $P(r,t)$. Initially, the probability density is spread out between $r\approx 0.5a_{\mu}$ and $r\approx 1.3a_{\mu}$ reflecting the repulsive interaction. After the system has evolved for $\approx \pi/2\omega$ the probability density is concentrated around $r=0$ like a bound state. The system varies between these two extremes with an approximate period of $\pi/\omega$.

\section{Conclusion}
\label{sec:Conc}

In this paper we consider the post-quench evolution of a harmonically trapped system of two contact-interacting bodies with the aim of determining if divergences in particle separation are present or not. We find that when quenching from finite $a_s$ to the non-interacting limit the particle separation (and thus total energy) diverges. This builds upon previous work that found the system size diverges when quenching from unitarity to the non-interacting limit \cite{kerin2020two, kerin2022quench, kerin2022effects}. Importantly, we observe that this divergent behaviour is present in \textit{only} these two cases; unitarity to the non-interacting regime and finite $a_s$ to the non-interacting regime. The divergence is not present in any other quench. Using the dynamic sweep theorem we are able to determine that the divergence arises due to the instantaneous nature of the quench rather than the zero-range nature of the interaction. We also calculate the evolution of the particle separation probability distribution and find a $1/r^2$ tail in $P(r,t\neq n\omega/\pi)$. 

Finally, we re-emphasize that these predictions of $P(r,t)$ are experimentally measurable. Modern techniques allow for low atom number systems to be constructed with high fidelity, \cite{PhysRevLett.96.030401, grunzweig2010near, serwane2011deterministic, zurn2012fermionization, wenz2013few, zurn2013pairing, murmann2015two} and Feshbach resonances are well understood \cite{fano1935feshbackh, feshbach1958feshbackh, tiesinga1993feshbackh, chin2010feshbach} and have been implemented in experiment before \cite{cetina2016ultrafast, skou2021non}. In particular, the evolution of the post-quench particle separation of a two atom system has been measured before \cite{guan2019density}, albeit the quench was in trap geometry not $a_s$. While particle separation will not diverge in experiments where $a_s$ is quenched to zero it is likely that oscillations in system size will still be large.

\section{Acknowledgements}

This research was supported by The University of Melbourne’s Research Computing Services and the Petascale Campus Initiative.

\bibliographystyle{apsrev4-1}{}
\bibliography{Few-Body-Refs}

\end{document}